\documentclass[aps,prl,twocolumn,showpacs,preprintnumbers,amsmath,amssymb,showpacs,showkeys]{revtex4}
\usepackage{amssymb}
\usepackage{amsmath}
\usepackage{amsfonts}
\usepackage{epsfig}
\usepackage{graphicx}
\usepackage{tabularx}

\newcommand{\seq}{\begin{subequations}}
\newcommand{\sen}{\end{subequations}}
\newcommand{\eq}{\begin{eqnarray}}
\newcommand{\en}{\end{eqnarray}}

\def\shiftdown#1{#1\llap{\lower.04ex\hbox{#1}}}

\newcommand{\ra}{\rangle}
\newcommand{\la}{\langle}

\newcommand{\bfb}{{\bf b}_{\perp}}

\def\arraystretch{1.5}

\begin{document}

\title{Chiral Symmetry Breaking and Meson Wave Functions in Soft-Wall AdS/QCD} 

\author{
Thomas Gutsche$^1$,
Valery E. Lyubovitskij$^1$
\footnote{On leave of absence
from Department of Physics, Tomsk State University,
634050 Tomsk, Russia},
Ivan Schmidt$^2$,
Alfredo Vega$^3$
\vspace*{.6\baselineskip}\\
}

\affiliation{
$^1$ Institut f\"ur Theoretische Physik,
Universit\"at T\"ubingen, \\
Kepler Center for Astro and Particle Physics,
\\ Auf der Morgenstelle 14, D-72076 T\"ubingen, Germany
\vspace*{.6\baselineskip} \\
\hspace*{-1cm}
$^2$ Departamento de F\'\i sica y Centro Cient\'\i
fico Tecnol\'ogico de Valpara\'\i so (CCTVal), Universidad T\'ecnica
Federico Santa Mar\'\i a, Casilla 110-V, Valpara\'\i so, Chile
\vspace*{.6\baselineskip} \\ 
$^3$ Departamento de F\'isica y Astronom\'ia, \\
Universidad de Valpara\'iso,\\
Avenida Gran Breta\~na 1111, Valpara\'iso, Chile
\vspace*{.5\baselineskip} \\ 
}

\date{\today}

\begin{abstract}

We consider mesons composed of light and heavy quarks
and discuss the construction of the corresponding meson 
wave functions in soft-wall AdS/QCD. We specifically take 
care that constraints imposed by chiral symmetry breaking 
and by the heavy quark limit are fulfilled.
The main results are 
i) the wave functions of light mesons 
have a nontrivial dependence on the current quark mass, 
which gives rise to a mass spectrum consistent with the 
one including explicit breaking of chiral symmetry; 
ii) the wave functions of heavy-light mesons 
generate their correct mass spectrum, the mass splittings 
of vector and pseudoscalar states, and the correct scaling
of leptonic decay constants $f_{Q\bar q} \sim 1/\sqrt{m_Q}$; 
iii) the wave functions of heavy quarkonia 
produce their correct mass spectrum and lead to 
a scaling behavior of the leptonic decay constants 
$f_{Q\bar Q} \sim \sqrt{m_Q}$ and 
$f_{c\bar b} \sim m_c/\sqrt{m_b}$ at $m_c \ll m_b$, 
consistent with potential models and QCD sum rules. 

\end{abstract}



\maketitle

The last decade has been marked by significant prog\-ress in 
the development of
AdS/QCD -- a new class of approaches based on gauge/gravity 
duality~\cite{Maldacena}, which try to model strong 
interactions in terms of fields propagating 
in extradimensional curved manifolds, e.g. in  
anti-de Sitter (AdS) space. Fields in AdS$_{d+1}$ space 
are classified by unitary, irreducible representations 
of the $SO(d,2)$ group~\cite{Flato,Ferrara}, 
which is isomorphic to the conformal group acting on the 
boundary of AdS space, where the dual conformal 
field theory (CFT) is living. With some assumptions the CFT 
can be truncated to QCD, and therefore the AdS fields 
could be holographically matched to QCD operators and bound states. 
There are two main types of AdS/QCD approaches: 
top-down (brane constructions in string theories leading to 
low-energy gauge theories with properties of QCD)
and bottom-up models (phenomenological 
frameworks specifying the geometry of AdS space and bulk fields 
in order to incorporate the basic properties of QCD). In this paper 
we focus on one of the successful examples of bottom-up approaches --- 
the soft-wall model (see e.g. Refs.~\cite{SW1,SW2,SW3}), which is based on  
a soft breaking of conformal invariance via the introduction of a dilaton 
field (holographic analogue of the gluon condensate~\cite{Aharony:1999ti})  
in the exponential prefactor or the effective potential. It leads to 
a truncation of AdS space in the infrared and therefore provides 
confinement of bulk fields, which are expanded in a tower of massive 
Kaluza-Klein modes identified with radial excitations of hadrons. 
The dilaton field is also responsible for the mechanism of spontaneous 
breaking of chiral symmetry. 

The profiles of bulk 
fields in extra dimension are matched to hadronic wave functions. 
It has been shown that the extradimensional coordinate can be 
identified with the transverse impact variable characterizing the separation 
of partons in a hadron in light-front QCD (Light-Front Holography)~\cite{LFH}.
It is essential to use the quadratic profile for the dilaton field to 
generate Regge trajectories for hadron masses and to reproduce 
the correct scaling of hadronic form factors at large values of 
Euclidean transverse momentum squared. It is also helpful to use
this profile since most of the calculations can be performed
analytically. Of course, not all features 
of QCD have been yet incorporated in the formalism of the soft-wall model. 
The work on providing an accurate matching to QCD is in progress. 
Therefore, some objections concerning the soft-wall model seem premature.

This work is addressed to the problem of constructing the 
hadronic wave functions using the AdS and light-front QCD correspondence. 
Originally the idea of such correspondence was proposed 
in Refs.~\cite{Brodsky:2007hb}. 
It was shown that, from the matching of matrix elements for physical 
processes (e.g. from the electromagnetic or gravitational form factors of 
hadrons), one can relate the string mode -- the bulk profile of the AdS field 
in a holographic dimension, and the transverse part of the hadronic 
light-front wave function (LFWF) for the case of massless quarks. 
Later in Ref.~\cite{Brodsky:2008pg}, in the case of a two-parton state the 
LFWF was generalized by the explicit inclusion of the constituent quark masses 
in the LF kinetic energy 
$\sum_i ({\bf k}_{\perp i}^2 + m_i^2)/x_i$. 
In the LFWF this corresponds to the introduction of the longitudinal 
WF corresponding to the so-called Brodsky-Huang-Lepage (BHL) or {\it Gaussian 
ansatz}~\cite{Brodsky:1982nx} (see also discussion 
in Ref.~\cite{Huang:1994dy}). 
In Refs.~\cite{SW5,SW5c} we studied the problem of the longitudinal 
part of the LFWF. In particular, in~\cite{SW5} following the ideas of 
Ref.~~\cite{Brodsky:2008pg}, we derived the longitudinal part of the LFWF 
using constraints of heavy quark effective theory. 
It was based 
on the BHL ansatz, where the dimensional parameter depends on the flavor 
of the constituent quarks.  
Note that the BHL ansatz deals with constituent quarks while the 
direct parameters of QCD are the current quarks. The main objective 
of this paper is to construct the longitudinal part of the LFWF in terms 
of current quark masses instead of constituent ones. In particular, 
we will show that this construction helps to introduce the mechanism 
of explicit breaking of chiral symmetry. 
This means that the explicit 
breaking of chiral symmetry is a property of the longitudinal part of 
the hadronic LFWF. The idea that explicit breaking of chiral 
symmetry is induced via the current quark mass dependence of the longitudinal 
LWFW was proposed in two-dimensional 
large $N_c$ QCD~\cite{'tHooft:1974hx}. This mechanism was later used
in the context of the two-dimensional massive Schwinger 
model~\cite{Bergknoff:1976xr,Ma:1987wi,Mo:1992sv}. 
Recently this problem was reexamined in 
Refs.~\cite{Chabysheva:2012fe,Forshaw:2012im}. 

In our approach the breaking of the conformal and 
chiral symmetries is related to the presence of the dilaton field. 
We show that both symmetries are broken dynamically due to the interaction 
with a dilaton -- background field living in the $z$ direction 
of AdS space; in other words, it does not show up at the AdS boundary. 
The dilaton is a massless zero mode. As it was mentioned before we use 
the quadratic dilaton profile, which can be considered as the expectation 
value of the scalar bulk field with dimension $2$, which is holographically 
dual to the dimension-2 gluon operator 
${\cal O}_{A^2} = \la A_{\mu, {\rm min}}^2\ra$~\cite{Gubarev:2000eu}. 
In different types of AdS/QCD approaches a few scenarios of chiral symmetry 
breaking have been proposed due to the presence of specific background 
or scalar fields. These fields can be considered as duals 
to the dimension-4 gluon ${\cal O}_{G^2} = {\rm tr}(G_{\mu\nu}^2)$ and 
the dimension-3 quark $O_{\bar qq} = \bar q q$ operators.
Their couplings to the bulk fields integrated over $z$ 
can be interpreted as holographic analogues of in-hadron condensates 
of the operators ${\cal O}_{A^2}$, ${\cal O}_{G^2}$ and ${\cal O}_{\bar qq}$. 
The reason for this assignment is the following: in AdS/QCD 
(especially in the soft-wall model) matrix elements are 
defined by integrals over profiles of bulk fields in the $z$ direction, 
which are holographic images of the hadronic wave functions. The 
idea of in-hadron condensates was suggested in~\cite{Casher:1974xd}
and developed in~\cite{Brodsky:2008xm,Brodsky:2010xf}. 
Notice that in-hadron condensates can be related to 
vacuum condensates (e.g. in the pion case via current algebra), and 
therefore there is no contradiction or conflict with QCD and chiral 
perturbation theory~\cite{Weinberg:1978kz}. In particular, the main idea 
of Refs.~\cite{Casher:1974xd,Brodsky:2008xm,Brodsky:2010xf} is that 
the quark and gluon condensates, holographically interpreted as 
spatial effects of bulk fields in the $z$ direction,
can be interpreted according to the
AdS/QCD dictionary as finite-volume effects in hadrons. 
As stressed in Ref.~\cite{Brodsky:2008xm} the quark and gluon condensates 
have spatial support within hadrons. 

We will show that our approach is consistent with model-independent 
relations and constraints valid in the regime of explicitly and 
spontaneously broken chiral symmetry. 
The pion is massless only when the chiral 
symmetry is spontaneously broken (for a vanishing current quark mass 
$\hat{m}$) while it becomes massive via the mechanism of explicit 
breaking of chiral symmetry encoded in the longitudinal part of its 
LFWF. In addition, we generate a finite splitting of the axial-vector 
and vector meson masses and reproduce the Weinberg 
relations between the masses of $\rho(770)$, $a_1(1270)$ and $f_0(600)$. 
When the current quark masses are included in 
the formalism through the longitudinal LFWF, the pion and other chiral 
pseudoscalar fields (kaon and $\eta$ meson) acquire a mass according 
to the Gell-Mann-Okubo-Renner relations. 

We would also like to mention that the classification of the 
bulk fields propagating in AdS$_5$ is similar to the classification 
of hadrons according to the chiral 
group~\cite{Glozman:2003bt,Glozman:2009bt}. 
As we stressed before, fields in AdS$_5$ are classified by unitary, 
irreducible representations of the $SO(4,2)$ group~\cite{Flato,Ferrara}. 
The $SO(4,2)$ group is decomposed with respect to 
its maximal subgroup $SO(4) \otimes SU(2)$, where  
$SO(4)$ is isomorphic to the group $SU(2) \otimes SU(2)$. 
Thus, the bulk fields are characterized by three quantum numbers ---  
minimal energy $E_0$ and two spins $J_1$ and $J_2$ --- 
and belong to the representations denoted by 
$D(E_0,J_1,J_2)$, which have specific chiral properties, because 
$SO(4)$ is also isomorphic to the chiral group 
$SU_L(2) \otimes SU_R(2)$. 
Because of the gauge/gravity duality the energy of the bulk 
field $E_0$ is identified with $\Delta$ --- the dimension of the corresponding 
CFT operator. Masses of bulk fields are expressed in terms of their energy. 
In particular, the scalar fields $S$ belong to 
the representation $D(E_0,0,0)$ with mass $\mu^2 R^2 = E_0 (E_0 - 4)$,
which also belongs to the chiral representation $(0, 0)$ (here $R$ is the AdS 
radius). There are two independent possibilities for a description of vector 
fields: in terms of vectors $V_\mu$ transforming according to the 
``vectorial'' representation $D(E_0, \frac{1}{2}, \frac{1}{2})$ with mass 
$\mu^2 R^2 = (E_0 - 1) (E_0 - 3)$, 
which belong to the chiral representation $(1, 0) \oplus (0, 1)$,
and in terms of antisymmetric tensors $W_{\mu\nu}$ transforming according to 
the ``tensorial'' representation $D(E_0, 1, 0) \oplus D(E_0, 0, 1)$ 
with mass $\mu^2 R^2 = (E_0 - 2)^2$, which belong to the chiral 
representation $(\frac{1}{2}, \frac{1}{2})$. Finally, the fermions $\psi$ 
with spin $1/2$ belong to the representation 
$D(E_0, 0, \frac{1}{2}) \oplus D(E_0, \frac{1}{2}, 0)$ 
with mass $\mu R = E_0 - 2$. Fermions belong to the chiral representation 
$(0, \frac{1}{2}) \oplus (\frac{1}{2}, 0)$. 
The classification of chiral properties of bulk fields in AdS is identical 
to the classification according to a chiral 
basis~\cite{Glozman:2003bt,Glozman:2009bt}, including 
two possible representations for vector fields, which correspond 
to vector and pseudotensor operators in QCD. In this sense, the 
case of exact chiral symmetry corresponds to gauge invariance for massless 
bulk fields in AdS. The chiral limit 
for vector mesons corresponds to conservation of the interpolating vector 
currents on the AdS boundary of AdS space. This means that their 
holographic analogues in AdS -- vector bulk fields, should be massless 
or have energy $E_0 = 2 + J = 3$ (with $J=1$) for both representations 
$D(E_0, \frac{1}{2}, \frac{1}{2})$ and 
$D(E_0, 1, 0) \oplus D(E_0, 0, 1)$.  Therefore in the case of exact chiral 
symmetry the twist dimension of interpolating operators of vector mesons 
must be $\Delta = 2 + J = 3$ (with $J=1$). 
However, this is only true for a conformal field theory. In order to guarantee 
the correct scaling of hadronic form factors in QCD the assignment 
$E_0 \equiv \Delta$ 
must be connected with the twist $\tau$ of the corresponding interpolating 
operator expressed in terms of $L = {\rm max} \, | L_z |$ --- the maximal
value of the $z$ component of the quark orbital angular momentum
in the LF wave function~\cite{SW2,SW4a}. 
In the case of two-parton states (mesons)  
$\Delta_M = \tau_M = 2 + L$, while in the case of three-parton states 
(baryons) $\Delta_B = \tau_B + 1/2 = 7/2 + L$. This means that the scaling 
of operators in QCD depending on $L$ corresponds to the picture of 
both broken conformal and chiral symmetries, and hadrons can be classified 
according to the nonrelativistic scheme $^{2S+1}L_J$. Conformal invariance is 
broken to the symmetry of the Poincar\'e group due to the
dilaton field present (living) in the extra $z$ dimension, 
which is dual to the gluon condensate~\cite{Aharony:1999ti}. 
The chiral group $SU(2)_L \otimes SU(2)_R$ 
is broken to the vector isospin group $SU(2)_V$ due to the coupling of bulk 
fields with the dilaton. 

In a series of papers two versions of soft-wall models 
based on the use of positive~\cite{SW2,SW4a,SW4b} and 
negative~\cite{SW5,SW6} dilaton profiles were developed. We showed 
that these models are equivalent in the case of the bound state problem due 
to a specific dilaton field-dependent redefinition of the bulk field. 
It could be shown that 
after such a redefinition the action for the bulk field  
$\Phi_J = \Phi_{M_1 \cdots M_J}(x,z)$ with spin $J$ 
reads~\cite{SW2,SW4a,SW4b,SW5,SW6}  
\eq\label{action_S}
\hspace*{-.25cm} 
S_J = \int d^d x dz \sqrt{g}
\Big[ \partial_M\Phi_J \, \partial^M\Phi^J  
- (\mu_J^2 + V_J(z)) \Phi_J \Phi^J \Big]\,, 
\en 
where the AdS metric is specified as 
$ds^2 = e^{2A(z)} (dx_\mu dx^\mu \\ 
- dz^2)$, $g = e^{5A(z)}$, 
$A(z)=\log(R/z)$ and $R$ is the AdS radius.  
Here $\Phi_J$ is the symmetric, traceless tensor 
classified by the representation $D(E_0,J/2,J/2)$ with 
energy $E_0 = \Delta = 2 + L$, which is related to the bulk 
mass $\mu_J$ as $\mu_J^2 R^2 = (E_0 - J) (E_0 - 4 + J) = L^2 - (2-J)^2$. 
This action is most convenient in order to study the bound 
state problem. Versions of the soft-wall model with so-called positive 
or negative dilaton profiles are just different manifestations of the 
action~(\ref{action_S}) after an appropriate dilaton-dependent redefinition 
of the bulk field $\Phi_J \to \Phi_J e^{\pm \varphi(z)/2}$.    
$V_J(z)$ is the effective dilaton potential, which has an analytical 
expression in terms of the field $\varphi(z)$ and the "metric" field $A(z)$ 
without referring to any specific form of their $z$ profiles:  
\eq 
V_J(z) = e^{-2A(z)} \Big( \varphi^{\prime\prime}(z)
+ (d-1-2J) \, \varphi^\prime(z) \, A^\prime(z) \Big) \; .
\en 
The potential $V_J(z)$ breaks both conformal and chiral invariance 
spontaneously. 
It was originally suggested in~\cite{SW1} to use a quadratic profile for 
the dilaton field (more precisely its $z$ profile) 
$\varphi(z) = \kappa^2 z^2$, where $\kappa$ is a
scale parameter, and the conformal metric $A(z) = \log(R/z)$ 
in order to obtain Regge behavior for the hadronic mass spectra. 
Note that such a form of the profile arises immediately if one considers 
the free action for the dilaton in the following form: 
\eq 
\hspace*{-.2cm} 
S_\chi = \int d^d x dz \sqrt{g}
\Big[ \partial_M\chi(x,z) \, \partial^M\chi(x,z) 
- \mu^2_\chi \chi^2(x,z) \Big] \,, 
\en 
where $\mu^2_\chi = \Delta (\Delta - 4) = - 4$ is the bulk mass of the 
dilaton with $\Delta = 2$ 
(therefore, the dilaton field is the pure scalar field). 
We propose that the Kaluza-Klein expansion for the dilaton field is trivial: 
\eq 
\chi(x,z) =  \varphi(z) \,,   
\en  
where $\varphi(z)$ 
is the dilaton $z$ profile, and therefore
the dilaton is only living in the $z$ direction. 
The dilaton should be massless (it does not appear in the observable hadronic 
mass spectra). Solving the equation of motion for
$\varphi(z)$ with $\mu^2_\chi R^2 = - 4$ and $A(z) = \log(R/z)$ and 
assuming the power behavior of $\varphi(z)$, we find $\varphi(z) \sim z^2$ 
which confirms the conjecture of Ref.~\cite{SW1}. 

As we stressed before, 
the quadratic dilaton can be considered as the expectation 
value of the scalar bulk field with dimension $2$, which is holographically 
dual to the dimension-2 gluon operator 
$\la A_{\mu, {\rm min}}^2\ra$~\cite{Gubarev:2000eu}  ---  
a nonlocal gauge-invariant gluon condensate coinciding with 
the gauge-noninvariant gluon condensate of the same dimension 
$\la A_{\mu}^2\ra$ in the Landau gauge. The dimension-2 gluon condensate 
has been discussed in the literature (see e.g. 
Refs.~\cite{SW3,Gubarev:2000eu,Celenza:1986th}).   
Note, the interpretation of the dilaton as the quantity dual to the condensate 
of the dimension-2 operator has been given in the framework of the soft-wall 
model~\cite{SW3} where the dilaton was introduced in the warping 
factor, breaking the conformal-invariant background metric. 
Inclusion of a more complicated form of the dilaton potential (e.g. taking 
into account self-interaction terms) can be viewed as a further extension 
of the soft-wall models based on the quadratic dilaton profile. 

Notice that a quadratic form of the $z$ profile of the 
dilaton is not unique. For example, in the Liu-Tseytlin 
model (a type of top-down AdS/QCD approach)~\cite{Liu:1999fc}
the conformal invariance was 
violated by the dilaton taken in the form $e^{\varphi(z)} = 1 + q z^4$, 
where the parameter $q$, according to the AdS/QCD 
dictionary~\cite{Klebanov:1999tb}, was related to 
the matrix element of a QCD operator:
the scalar $\la {\rm tr}(G_{\mu\nu}^2) \ra$ and pseudoscalar 
$\la {\rm tr}(G_{\mu\nu} \tilde G_{\mu\nu})\ra$ gluon condensates. 
On the other hand, the scalar dimension-4 gluon condensate 
is connected to the quark vacuum condensate via the decoupling 
relation~\cite{Shifman:1978bx} 
\eq 
\la 0| \bar q q |0 \ra \simeq - \frac{1}{12 m} 
\la 0| \frac{\alpha_s}{\pi} \ {\rm tr}(G_{\mu\nu}^2)|0 \ra 
\en  
derived in the leading order of a $1/m$ expansion, where $m$ is
the mass of a heavy quark or the constituent mass in the case of 
light quarks. Therefore, nonzero dimension-4 gluon condensates 
signal the existence of nonzero dimension-6 quark condensates, 
which are manifestations of the spontaneous breaking of chiral 
symmetry. Therefore we see that the breaking of conformal symmetry 
is connected with the  breaking of chiral symmetry. 
Actually, this idea is not new (see e.g. discussion in 
Ref.~\cite{Crewther:1971bt}). 

Next we explain why the dilaton is also responsible for 
the spontaneous breaking of chiral symmetry.  
With the quadratic $z$ profile 
of the dilaton field $\varphi(z) = \kappa^2 z^2$
the meson spectrum is given by the master formula~\cite{SW5}  
\eq\label{Eq1} 
M^2_{nJ} = 4 \kappa^2 \biggl( n + \frac{L + J}{2}\biggr) \, 
\en 
where $\kappa$ is the dilaton parameter, which signals the spontaneous 
breaking of conformal and chiral symmetry and leads to a discrete 
mass spectrum of hadrons. An additional mechanism of 
spontaneous breaking of chiral symmetry is encoded in the bulk mass 
$\mu_J^2$,  which explicitly depends on $L$ and forbids parity doubling (e.g. 
between vector and axial mesons). The corresponding breaking term in the mass 
formula is $\delta M^2 = 2 \kappa^2 (L - J)$. It means that 
we can interpret Eq.~(\ref{Eq1}) 
as 
\eq 
M^2 = \bar{M}^2 + \delta M^2 
\en  
where $\bar{M}^2 = 4 \kappa^2 (n + J)$ is the term corresponding to 
the parity doubling limit. 

The hadronic wave functions are identified with the profiles of 
AdS modes $\Phi_{n}(z)$ in the $z$ direction: 
\eq
\hspace*{-.2cm}
\Phi_{nL}(z) = \sqrt{\frac{2 \Gamma(n+1)}{\Gamma(n+L+1)}} \kappa^{L+1}
z^{2+L} e^{-\kappa^2z^2/2}  
L_n^{L}(\kappa^2z^2)\,, 
\en 
which have the correct behavior in both the
ultraviolet and infrared limits, 
\eq
\Phi_{nL}(z) \to z^{2+L}  \ {\rm at \ small} \ z, \ 
\Phi_{nL}(z) \to 0 \  {\rm at \ large} \ z \,
\en 
and are normalized according to the condition 
\eq 
\int\limits_0^\infty \frac{dz}{z^3} 
\Phi_{nL}^2(z) = 
\int\limits_0^\infty dz
\phi_{nL}^2(z) = 1 \,, 
\en 
where $\Phi_{nL}(z) = z^{3/2} \, \phi_{nL}(z)$. 
At $z \to 0$ the scaling of the bulk profile is identified 
with the scaling of the corresponding mesonic interpolating operator 
$\tau = 2 + L$. As we mentioned in the Introduction, $\tau$ 
depends on $L$ (instead of $J$ as in CFT) because we model QCD and 
should reproduce the scaling of hadronic form factors. As we stressed before, 
dependence on $L$ means spontaneous breaking of chiral invariance, 
which is expected 
because after the introduction of the dilaton field we broke the conformal 
or gauge invariance acting in AdS space. As we noted 
before, the chiral group is isomorphic to the subgroup of $SO(4,2)$. 

Let us discuss the consistency of the mass formula~(\ref{Eq1}). 
First of all, we reproduce the massless pion $M_\pi^2 = 0$ 
for $n=L=J=0$, which is consistent with the picture of spontaneous breaking 
of chiral symmetry. 
Also, at $z=0$ all bulk profiles vanish, which corresponds to pointlike 
hadrons (zero scale limit) and we restore conformal,  
chiral, and gauge invariance associated with the AdS group $SO(4,2)$. 
In this sense we do not have any contradictions with chiral invariance 
as mentioned in Ref.~\cite{Glozman:2009bt}, because for $z=0$ it is 
restored for harmonics with any $L$ and for finite $z$ we live
in the phase of spontaneously broken chiral invariance (see further 
details in~\cite{deTeramond:2009qx}). 
For finite $z$ the profile for the ground state $\rho$ meson  
is defined by their leading twist corresponding to $L=0$, 
which is consistent with the statement of Ref.~\cite{Glozman:2009bt}, 
that the ground state $\rho$ meson is almost a pure $S$ wave 
in the infrared. On the other hand, the bulk profile dual to the ground 
state $\rho$ meson is defined in the extra dimension. 
Therefore it has no correspondence to the chiral-invariant 
superposition of Fock states with $L=0$ ($S$-wave) and $L=2$ ($D$-wave) 
discussed in~\cite{Glozman:2009bt}.   
Note that both the electromagnetic field and $J=L=1$ vector mesons 
have the same interpolating twist-3 operator. They also have
the same dual field in AdS space --- a massless vector bulk field and 
the bulk-to-boundary propagator of the electromagnetic field 
in the timelike region~\cite{Grigoryan:2007my,SW4a}, which 
has the poles corresponding to the excited states with $J=L=1$: 
\eq\label{Eq2}
p^2 = M^2 = 4 \kappa^2 (n + 1) 
\en 
as expected in the chiral-invariant limit. 

Let us mention other interesting results deduced from the  
meson mass formula~(\ref{Eq1}). There are two relations 
between the masses of the ground states of vector $\rho(770)$, 
axial-vector $a_1(1270)$, and scalar 
$f_0(600)$ mesons,  
\eq 
M_{a_1} = M_\rho \, \sqrt{2} = 2 \kappa \,, \quad\quad 
M_{f_0} = M_\rho = \sqrt{2}\, \kappa   
\en     
which are consistent with the predictions done in 
Ref.~\cite{Weinberg:1990xn} in the same limit when chiral 
symmetry is spontaneously broken. 
Also, the vector and axial-vector multiplets are not degenerate in 
our approach (even for higher values of $J$), because of the finite 
mass splitting of axial-vector and vector mesons states: 
\eq 
\Delta_{_{VA}} = M_A^2 - M_V^2 = 2 \kappa^2 \,. 
\en
This means that we do not have parity doubling. 
An amazing fact is that the same prediction 
relating the masses of $\rho$ and $a_1$ mesons was also 
obtained~\cite{Weinberg:1967kj} on 
the basis of spectral function sum rules at $M_\pi = 0$ and 
using the Kawarabayashi-Suzuki-Riazuddin-Fayyazuddin 
formula~\cite{Kawarabayashi:1966kd} for the $\rho\pi\pi$ coupling. 

For the value $\kappa = 500$ MeV used in this paper 
the masses of the $\rho$ and $a_1$ mesons 
are well reproduced in comparison to data 
\eq 
\hspace*{-.2cm} 
M_\rho  &=& 721 \ {\rm MeV}\,, \ ({\rm data:} \  775.49 \pm 0.34 
\ {\rm MeV}) \,,
\nonumber\\
\hspace*{-.2cm} 
M_{a_1} &=& 1010  \ {\rm MeV}\,, \ ({\rm data:} \  1230 \pm 40 
\ {\rm MeV}) 
\,,
\en 
while our result for the $f_0$ mass should be considered 
as a prediction. We get $M_{f_0} = 721$ MeV, which 
perfectly agrees with a model-independent 
result based on analyticity and unitarity of 
the $S$ matrix~\cite{Surovtsev:2011yg}: 
$M_{f_0} = 735.0 \pm 6.1$~MeV. 
But also note that a recent compilation~\cite{Beringer:1900zz}
indicates a mass of $M_{f_0} = 446 \pm 6$ MeV as deduced from the
pole position in the process amplitude.

Next we consider the inclusion of the longitudinal part of the LFWF. 
We will show that this extension is important to include in our 
formalism the dependence on the current quark masses and therefore 
explicit breaking of chiral symmetry. 
As we stressed before, the first step in this 
direction was done in Refs.~\cite{Brodsky:2008pg}. The authors
proposed to write down the mesonic two-parton wave function 
in a factorized form, as a product of transverse $\phi_{nL}(\zeta)$, 
longitudinal $f(x,m_1,m_2)$ and angular $e^{im\phi}$
modes. In Ref.~\cite{SW5} we further proposed to do such separation in 
a more convenient form, factorizing in addition $\sqrt{x (1-x)}$ --- 
the Jacobian of the $\zeta \to |\bfb|$ coordinate transformation: 
\eq
\psi_{q_1\bar q_2}(x,\zeta,m_1,m_2) &=& 
\frac{\phi_{nL}(\zeta)}{\sqrt{2\pi\zeta}}
f(x,m_1,m_2) \nonumber\\
&\times& e^{im\phi} \, \sqrt{x(1-x)} \,. 
\en 
In Ref.~\cite{SW5} we used a {\it Gaussian ansatz} for the 
longitudinal part of the LFWF and treated the quark masses $m_1$ and $m_2$ 
as constituent quark masses. Here we follow 
Refs.~\cite{'tHooft:1974hx,Bergknoff:1976xr} and consider current 
quark masses, and by an appropriate choice of the longitudinal wave function 
$f(x,m_1,m_2)$ we get consistency with QCD in both sectors of light 
and heavy quarks. We generate the masses of light 
pseudoscalar mesons in agreement with the picture resulting from 
explicit breaking of chiral symmetry --- in the leading order of 
the chiral expansion the masses of pseudoscalar mesons are linear 
in the current quark mass. In this vein we also guarantee that the 
pseudoscalar meson masses satisfy the Gell-Mann-Oakes-Renner (GMOR) 
relation for the pion mass 
\eq 
M_\pi^2 = 2 \hat{m} \, B 
\en 
and the Gell-Mann-Okubo (GMO) relation between the masses of pion, kaon,  
and $\eta$ meson: 
\eq 
4 M_K^2 \,=\, M_\pi^2 \, + \, 3 M_\eta^2 \,, 
\en
In previous equations $\hat{m} = (m_u + m_d)/2$ is the average mass 
of $u$ and $d$ quarks (we work in the isospin limit $m_u = m_d$), 
$B = | \la 0 | \bar u u | 0 \ra |/F_\pi^2$ is the quark condensate 
parameter, and $F_\pi$ is the leptonic decay constant. 
Note, the condensate parameter $B$ is related to 
the coupling constant of the pseudoscalar density to the 
pion $G_\pi$~\cite{Weinberg:1978kz} (or in-hadron condensate of 
pion~\cite{Brodsky:2008xm,Brodsky:2010xf}): 
\eq
B = \frac{G_\pi}{2F_\pi}\,, \quad 
\la 0| \bar q i\gamma_5 \tau^i q | \pi^k \ra = \delta^{ik} \, G_\pi \,. 
\en  
In the sector of heavy quarks we get agreement with heavy quark effective 
theory and potential models for heavy quarkonia.  
In the heavy quark mass limit 
$m_Q \to \infty$ we obtain the correct scaling of 
the leptonic decay constants for both heavy-light mesons 
$f_{Q\bar q} \sim 1/\sqrt{m_Q}$ and heavy quarkonia  
$f_{Q\bar Q} \sim \sqrt{m_Q}$ and 
$f_{c\bar b} \sim m_c/\sqrt{m_b}$ at $m_c \ll m_b$. 
In this limit we also generate the correct expansion of heavy meson masses 
\eq 
M_{Q\bar q} &=& m_Q + \bar\Lambda + {\cal O}(1/m_Q) \,, \nonumber\\
M_{Q\bar Q} &=& 2 m_Q + E + {\cal O}(1/m_Q) \,, 
\en 
where $\bar\Lambda$ is the approximate difference between the masses 
of the heavy-light meson and the heavy quark, $E$ is the binding energy 
in heavy quarkonia, and their splittings, e.g. between vector and 
pseudoscalar states of heavy-light mesons, become  
\eq 
M_{Q\bar q}^V - M_{Q\bar q}^P \, \sim \, \frac{1}{m_Q} \,. 
\en  
We choose the longitudinal wave function in the form 
\eq\label{LLWF} 
f(x,m_1,m_2) = N \, x^{\alpha_1} \, (1-x)^{\alpha_2} \, 
\en 
where $N$ is the normalization constant fixed from
\eq
1 = \int\limits_0^1 dx \, f^2(x,m_1,m_2) 
\en
and $\alpha_1, \alpha_2$ are parameters that 
will be fixed in order to get consistency with QCD. 

In the present paper the physical quantities of interest are the 
mass spectrum $M^2_{nJ}$ and lepton decay constants $f_{M}$ of mesons, 
which are given by the expressions~\cite{SW5}  
\eq\label{M2_mesons} 
M^2_{nJ} &=& 4 \kappa^2 \biggl( n + \frac{L + J}{2} \biggr)\nonumber\\
&+& \int\limits_0^1 dx
\biggl( \frac{m_1^2}{x} + \frac{m_2^2}{1-x} \biggr) f^2(x,m_1,m_2)\,,\\
f_M 
&=& \kappa \frac{\sqrt{6}}{\pi} \int\limits_0^1dx \sqrt{x(1-x)} \,
 f(x,m_1,m_2) \,. 
\en
Using our ansatz for the longitudinal wave function~(\ref{LLWF}) 
we calculate the leptonic decay constant and the correction to 
the mass spectrum 
analytically~\cite{SW5}: 
\eq\label{M2_mesons_results} 
M^2_{nJ} &=& 4 \kappa^2 \biggl( n + \frac{L + J}{2} \biggr)\nonumber\\
         &+& (1 + 2 \alpha_1 + 2 \alpha_2) \, 
\biggl( \frac{m_1^2}{2\alpha_1} 
      + \frac{m_2^2}{2\alpha_2} 
\biggr)\,, \nonumber\\
f_M &=& \kappa \, \frac{\sqrt{6}}{\pi} \, 
\frac{\Gamma(\frac{3}{2} + \alpha_1) \, \Gamma(\frac{3}{2} + \alpha_2)} 
{\Gamma(3 + \alpha_1 + \alpha_2)} \nonumber\\
&\times&\sqrt{\frac{\Gamma(2 + 2\alpha_1 + 2\alpha_2)}
{\Gamma(1 + 2\alpha_1)\,\Gamma(1 + 2\alpha_2)}} \,. 
\en
Next we analyze different types of mesons. 
We start with the light pseudoscalar mesons. Here we want to 
incorporate a mechanism for explicit breaking 
of chiral symmetry in order to reproduce the GMOR and GMO relations 
in the leading order of the chiral expansion. For this purpose we set 
$\alpha_i = m_i/(2 B)$, which is consistent with the ideas and results of 
Refs.~\cite{'tHooft:1974hx,Bergknoff:1976xr} 
and which means that $\alpha_i$ vanishes in the chiral limit $m_i \to 0$.  
The leptonic decay constants of both the pion and kaon are finite 
and degenerate in the chiral limit~\cite{Brodsky:2007hb,SW5}:
\eq
f_P = \frac{\kappa \sqrt6}{8}\,.
\en
Concerning the longitudinal wave functions of other light mesons, 
the choice of the $\alpha_i$ parameters does not necessarily coincide with the
one used for the pseu\-do\-sca\-lar mesons.
But for simplicity we will 
use the same universal longitudinal wave function for all light mesons. 
In this case the leptonic decay constants of $\pi$, $K$, and $\rho^0$ 
mesons are degenerate and the corresponding constants for $\omega$ and $\phi$ 
mesons are related via the SU(3) flavor conditions: 
\eq 
f_\pi = f_K = f_\rho = 3 f_\omega 
      = \frac{3 f_\phi}{\sqrt{2}}= \kappa \frac{\sqrt{6}}{8} \,. 
\en 
Next we consider the heavy-light mesons. 
Here we should reproduce the heavy quark mass expansion of 
the heavy-light meson masses, the mass splitting of vector and pseudoscalar 
states, and the $1/\sqrt{m_Q}$ scaling of the leptonic decay constants. 
All these constraints are fulfilled if the $\alpha_i$ 
parameters are fixed as $\alpha_Q = \alpha$, a flavor-independent constant, 
while the parameter $\alpha_q$ must be 
\eq 
\alpha_q = \frac{2 \alpha_Q}{m_Q} \biggl( 1 + \frac{\bar\Lambda}{2m_Q}\biggr) 
- \frac{1}{2} \;.
\en  
For this choice the results for the mass spectrum and 
leptonic decay constants, in the leading order of the chiral expansion, and the
leading and next-to-leading 
order of the heavy quark mass expansion, are  
\eq 
M_{Q\bar q}^2 &=& 4 \kappa^2 \biggl( n + \frac{L + J}{2}\biggr) 
+ (m_Q + \bar\Lambda)^2 \,, \\
f_{Q\bar q} &=& \frac{\kappa \sqrt{6}}{\pi} \, 
\frac{2\sqrt{\alpha}}{\alpha + \frac{3}{2}} \, 
\sqrt{\frac{\bar\Lambda}{m_Q}} \,  
\sim \, \sqrt{\frac{1}{m_Q}}  \,. 
\en 
Finally, for heavy quarkonia we fix the parameters $\alpha_{Q_i}$ as 
\eq 
\alpha_{Q_i} = \frac{m_{Q_i}}{4 E} \, 
\biggl( 1 - \frac{E}{2 (m_{Q_1} + m_{Q_2})} \biggr) \,+\, 
{\cal O}\Big(\frac{1}{m_{Q_i}}\Big) 
\en 
and then we get the following results for the spectrum 
\eq 
M_{Q_1\bar Q_2}^2 = 4 \kappa^2 
\biggl( n + \frac{L + J}{2}\biggr) + (m_{Q_1} + m_{Q_2} + E)^2 \,. 
\en 
It was shown in~\cite{Sergeenko:1993sn} that the
trajectories of bottomia states deviate from linearity as we also
discussed in~\cite{SW5}. This effect can be related to 
the one--gluon exchange term, which results in an additional Coulomb--like
interaction between quarks $V(r) = - 4\alpha_s/3r$, where $\alpha_s$
is the strong coupling constant. Its contribution to the mass
spectrum $M^2$ is negative and proportional to the quark mass
squared~\cite{Sergeenko:1993sn}. For
light and heavy-light mesons this term can 
be safely neglected, while this is not the case for heavy quarkonia 
(especially for bottomia states). 
Extending the results of Refs.~\cite{Sergeenko:1993sn} 
to the general case of a meson containing constituent quarks with masses 
$m_{Q_1}$ and $m_{Q_2}$, we
get the following expression for the shift of $M^2$ due to the color
Coulomb potential:
\eq
\Delta M^2_{Q_1Q_2} = 
- \frac{64\alpha_s^2m_{Q_1}m_{Q_2}}{9 \, (n+L+1)^2} \,,
\en
where $\alpha_s$ is the strong coupling considered as a free parameter. 
Therefore, the final expression for the heavy quarkonia spectra is 
given by the master formula: 
\eq 
M_{Q_1\bar Q_2}^2 &=& 4 \kappa^2 
\biggl( n + \frac{L + J}{2}\biggr) + (m_{Q_1} + m_{Q_2} + E)^2 
\nonumber\\
&-& \frac{64\alpha_s^2m_{Q_1}m_{Q_2}}{9 \, (n+L+1)^2} \,. 
\en 
For the leptonic decay constants of unflavored quarkonia 
we get the following result in leading order of the $1/m_Q$ expansion:
\eq\label{fQQ} 
f_{Q\bar Q} = \frac{\kappa \sqrt{6}}{(2\pi)^{3/4}} \, 
\biggl(\frac{E}{m_Q}\biggr)^{1/4}  
\en 
and for the decay constant of $B_c$ meson: 
\eq\label{fBc} 
f_{c\bar b} = \frac{2 \kappa \sqrt{6}}{\pi^{3/4}} \, 
\frac{(m_c/E)^{3/4}}{(m_b/E)} \,.  
\en 
In the case of the $B_c$ meson we additionally apply the 
condition $m_c \ll m_b$. 
Equations~(\ref{fQQ}) and (\ref{fBc}) can be combined into 
a general formula for the leptonic decay of heavy quarkonia: 
\eq 
f_{Q_1\bar Q_2} = \frac{2 \kappa \sqrt{6}}{\pi^{3/4}} \, 
\frac{(\mu_{Q_1Q_2}/E)^{3/4}}{(M_0/E)} \,,   
\en 
where $\mu_{Q_1Q_2} = m_{Q_1} m_{Q_2}/(m_{Q_1} + m_{Q_2})$ is 
the reduced mass of heavy quarkonia and $M_0 = m_{Q_1} + m_{Q_2}$.  
In order to get the correct scaling~\cite{Kiselev:1992au} 
of the leptonic decay constants 
of heavy quarkonia we propose the following: in the case of heavy quarkonia
the dilaton parameter $\kappa$ should be flavor dependent and scale 
as $\mu^{1/4}_{Q_1Q_2} \cdot M_0^{1/2}$. 
Note that different dilaton parameters for light mesons and 
heavy quarkonia were also considered before 
(see e.g. Ref.~\cite{Fujita:2009ca}). 
Here we use the following ansatz for $\kappa$: 
\eq 
\kappa = \beta \, \biggl( \frac{\mu_{Q_1Q_2}}{E} \biggr)^{1/4} \, 
                  \biggl( \frac{M_0}{E} \biggr)^{1/2} \,, 
\en 
where $\beta = {\cal O}(m_Q^0)$. 
For unflavored quarkonia the final result for the 
leptonic decay constant reads: 
\eq 
f_{Q\bar Q} = \beta \, \frac{\sqrt{3}}{\pi^{3/4}} \, 
\sqrt{\frac{m_Q}{E}}  \,. 
\en
For the $B_c$ meson we get 
\eq 
f_{c\bar b} = 2\beta \, \frac{\sqrt{6}}{\pi^{3/4}} \, 
\frac{m_c/E}{\sqrt{m_b/E}}  \,. 
\en 
Finally we present the numerical results. 
The parameters are fixed to the following values. For the
current quark masses we use 
\eq\label{const_qm}
& &m_u = m_d = \hat{m} = 7 \ {\rm MeV}\,, \nonumber\\
& &m_s = 24 \hat{m} = 168 \ {\rm MeV}\,, \nonumber\\
& &m_c = 1.275 \ {\rm GeV}\,, \hspace*{.5cm}
   m_b = 4.18  \ {\rm GeV}\,.
\en 
For the strong coupling constants $\alpha_s$ 
we use the following set of parameters:  
\eq 
\alpha_s(c\bar c) = 0.45\,, \hspace*{.2cm}
\alpha_s(c\bar b) = 0.383\,, \hspace*{.2cm}
\alpha_s(b\bar b) = 0.27 \,. 
\en 
Note that for the light and heavy-light mesons we use the universal
parameter $\kappa = 500$ MeV.  
The parameters $\bar\Lambda$ for heavy-light mesons are 
fixed from the mass difference of the experimental 
values of ground state of $D$, $D_s$, $B$, $B_s$ 
mesons and corresponding heavy quark mass as 
\eq 
& &\bar\Lambda_{cq} = 0.595 \ {\rm GeV}\,, \ 
   \bar\Lambda_{cs} = 0.695 \ {\rm GeV}\,, \nonumber\\
& &\bar\Lambda_{bq} = 1.1 \ {\rm GeV}\,, \ \ \ \ 
   \bar\Lambda_{bs} = 1.19 \ {\rm GeV}\,. 
\en 
The binding energies for heavy quarkonia are fixed as 
\eq 
& &E_{cc} = 0.795 \ {\rm GeV}\,, \hspace*{.2cm} 
E_{cb} = 1.25 \ {\rm GeV}\,, \nonumber\\ 
& &E_{bb} = 1.45 \ {\rm GeV}\,.
\en 
The set of $\beta$ couplings, defining the heavy flavor dependence of the  
dilaton parameter $\kappa$, is fixed in the case of heavy quarkonia as  
\eq
& &\beta(c\bar c) = 0.36 \ {\rm GeV}\,, \hspace*{.2cm} 
\beta(c\bar b) = 0.32 \ {\rm GeV}\,, \nonumber\\ 
& &\beta(b\bar b) = 0.41 \ {\rm GeV}\,.
\en 
These parameters are nearly the same for all 
quarkonia states, which is consistent with proposed scaling 
$\beta = {\cal O}(m_Q^0)$. 
In Tables~\ref{tab:1}-\ref{tab:6} we present the numerical results 
both for the mass spectrum and the leptonic decay constants of light, 
heavy-light and heavy quarkonia in leading order of the chiral and 
heavy-quark mass expansion. Results are for most of the part
in reasonable agreement with data.
We would like to stress that we significantly improved 
the description of mesonic properties in comparison to our previous
efforts~\cite{SW5}. 

In conclusion, we demonstrated that in the soft-wall mo\-del, where 
conformal or $SO(4,2)$ gauge invariance is broken the same applies for
chiral invariance which is also broken (spontaneously). 
This is manifested in the $L$ dependence of twists of interpolating 
operators of hadrons and in their observables
such as mass spectra and form factors. 
In the limit of chiral invariance  one could expect a dependence on $J$. 
However, the $J$ dependence is specific 
for conformal field theories and not for QCD. 
Namely, the $L$ dependence is dictated by the scaling of hadronic form 
factors at higher $Q^2$.A restriction to a specific $L$ 
(in our case to the minimal $L$ for a specific hadron) is a manifestation 
of spontaneous breaking of chiral symmetry. 
Chiral symmetry is restored 
in the exact limit $z=0$ (not for small $z$), which is trivial and 
corresponds to the restoration of chiral symmetry for all hadrons having 
components with an adjustable value of $L$. We demonstrated explicitly
that the present approach 
is consistent with model-independent predictions obtained in the case of 
spontaneous broken chiral symmetry: i) a massless pion, 
ii) the Weinberg sum rule relating masses of $\rho$, $a_1$ and $f_0$ mesons. 
In the chiral limit the hadron eigenstates are superpositions of components 
with different values of $L$. 
Finally, we demonstrated how to consistently construct the longitudinal LFWF 
and include the current quark mass dependence. The latter is consistent, 
in the light quark 
sector, with the mechanism of explicit breaking of chiral symmetry, 
and in the heavy quark sector with the heavy quark spin-flavor symmetry. 

\begin{acknowledgments} 

The authors thank Stan Brodsky and Guy de T\'e\-ra\-mond 
for useful discussions. 
This work was supported by the DFG under Contract No. LY 114/2-1,
by Federal Targeted Program ``Scientific
and scientific-pedagogical personnel of innovative Russia''
Contract No. 02.740.11.0238, by FONDECYT (Chile) under Grant No. 1100287 
and by CONICYT (Chile) under Grant No. 7912010025. 
The work is done partially under 
the project 2.3684.2011 of Tomsk State University. 
V.E.L. would like to thank Departamento de F\'\i sica y Centro
Cient\'\i fico Tecnol\'ogico de Valpara\'\i so (CCTVal), Universidad
T\'ecnica Federico Santa Mar\'\i a, Valpara\'\i so, Chile for warm
hospitality. 

\end{acknowledgments}

\newpage 

\begin{widetext}

\begin{table}
\caption{Masses of light mesons. \label{tab:1}}
\begin{tabular}{|l|c|c|c|l|l|l|l|}
\hline
Meson&$n$&$L$&$S$&\multicolumn{4}{c|}{Mass [MeV]} \\
\hline
$\pi$&0,1,2,3&0&0&$M_{\pi(140)}=140$&$M_{\pi(1300)}=1010$
&$M_{\pi(1800)}=1421$&$M_{\pi(4s)}=1738$ \\ \hline
$K$& 0&0,1,2,3&$\;0\;$&$M_{K}=495$&$ M_{K_1(1270)}=1116$
&$ M_{K_2(1770)}=1498$ & $M_{K_3}=1801$ \\ \hline
$\eta$&0,1,2,3&0&0&$M_{\eta(1s)}=566$
&$ M_{\eta(2s)}=1149$&$ M_{\eta(3s)}=1523$
&$ M_{\eta(4s)}=1822$ \\ \hline
$f_0[\bar n n]$&0,1,2,3&1&1&$M_{f_0(1p)}=721$&$M_{f_0(2p)}=1233$
&$M_{f_0(3p)}=1587$&$M_{f_0(4p)}=1876$ \\ \hline
$f_0[\bar s s]$&0,1,2,3&1&1&$M_{f_0(1p)}=985$&$M_{f_0(2p)}=1404$
&$M_{f_0(3p)}=1723$&$M_{f_0(4p)}=1993$ \\ \hline
$\rho(770)$&0,1,2,3&0&1&$M_{\rho(770)}=721$&$M_{\rho(1450)}=1233$
&$M_{\rho(1700)}=1587$&$M_{\rho(4s)}=1876$ \\ \hline
$\omega(782)$&0,1,2,3&0&1&$M_{\omega(782)}=721$&$M_{\omega(1420)}=1233$
&$M_{\omega(1650)}=1587$&$M_{\omega(4s)}=1876$ \\ \hline
$\phi(1020)$ &0,1,2,3&0&1&$M_{\phi(1s)}=985$&$M_{\phi(2s)}=1404$
&$M_{\phi(3s)}=1723$&$M_{\phi(4s)}=1993$ \\ \hline
$a_1(1260)$&0,1,2,3&1&1&$M_{a_1(1p)}=1010$&$M_{a_1(2p)}=1421$
&$M_{a_1(3p)}=1738$&$M_{a_1(4p)}=2005$ \\ \hline
\end{tabular}

\vspace*{.5cm}

\caption{Masses of heavy--light mesons. \label{tab:2}}
\begin{tabular}{|l|c|c|c|c|c|c|c|c|}
\hline
Meson&$J^{\rm P}$&$n$&$L$&$S$&\multicolumn{4}{c|}{Mass [MeV]} \\
\hline
$D(1870)$&$0^{-}$&0&0,1,2,3         &0& 1870 & 2000 & 2121 & 2235 \\ \hline
$D^{\ast}(2010)$&$1^{-}$&0&0,1,2,3  &1& 2000 & 2121 & 2235 & 2345 \\ \hline
$D_s(1969)$&$0^{-}$&0&0,1,2,3       &0& 1970 & 2093 & 2209 & 2320 \\ \hline
$D^{\ast}_s(2107)$&$1^{-}$&0&0,1,2,3&1& 2093 & 2209 & 2320 & 2425 \\ \hline
$B(5279)$&$0^{-}$&0&0,1,2,3         &0& 5280 & 5327 & 5374 & 5420 \\ \hline
$B^{\ast}(5325)$&$1^{-}$&0&0,1,2,3  &1& 5336 & 5374 & 5420 & 5466 \\ \hline
$B_s(5366)$&$0^{-}$&0&0,1,2,3       &0& 5370 & 5416 & 5462 & 5508 \\ \hline
$B^{\ast}_s(5413)$&$1^{-}$&0&0,1,2,3&1& 5416 & 5462 & 5508 & 5553 \\ \hline
\end{tabular}

\vspace*{.5cm}

\caption{Masses of heavy quarkonia $c\bar c$, $b\bar b$ and
$c \bar b$}
\label{tab:3}
\begin{tabular}{|l|c|c|c|c|l|l|l|l|}
\hline
Meson&$J^{\rm P}$&$n$&$L$&$S$&\multicolumn{4}{c|}{Mass [MeV]} \\ \hline
$\eta_c(2980)$&$0^{-}$ &0,1,2,3&0&0  & 2975 & 3477 & 3729 & 3938 \\\hline
$\psi(3097)$ &$1^{-}$ &0,1,2,3&0&1  & 3097 & 3583 & 3828 & 4032 \\ \hline
$\chi_{c0}(3415)$&$0^{+}$&0,1,2,3&1&1& 3369 & 3628 & 3843 & 4038 \\ \hline
$\chi_{c1}(3510)$&$1^{+}$&0,1,2,3&1&1& 3477 & 3729 & 3938 & 4129 \\ \hline
$\chi_{c2}(3555)$&$2^{+}$&0,1,2,3&1&1& 3583 & 3828 & 4032 & 4219 \\ \hline
$\eta_{b}(9390)$&$0^{-}$ &0,1,2,3&0&0& 9337 & 9931 & 10224 & 10471\\ \hline
$\Upsilon(9460)$&$1^{-}$ &0,1,2,3&0&1& 9460 & 10048& 10338& 10581 \\ \hline
$\chi_{b0}(9860)$&$0^{+}$&0,1,2,3&1&1& 9813 & 10110& 10359& 10591 \\ \hline
$\chi_{b1}(9893)$&$1^{+}$&0,1,2,3&1&1& 9931 & 10224& 10471& 10700 \\ \hline
$\chi_{b2}(9912)$&$2^{+}$&0,1,2,3&1&1&10048 & 10338& 10581& 10808 \\ \hline
$B_c(6277)$&$0^{-}$      &0,1,2,3&0&0& 6277 & 6719 & 6892 & 7025  \\ \hline
\end{tabular}
\end{table}

\begin{table}
\def\arraystretch{1.25}
\caption{Decay constants $f_P$ of pseudoscalar mesons in MeV.}
\label{tab:4}
\begin{tabular}{|l|c|c|}
\hline
Meson &Data~\cite{Beringer:1900zz} & \ Our \ \\ \hline
$\pi^-$& $130.4\pm 0.03 \pm 0.2$ & 153 \\
\hline
$K^-$ & $156.1\pm 0.2 \pm 0.8$ & 153 \\
\hline
$D^+$ & $206.7 \pm 8.9$ & 207 \\
\hline
$D_s^+$ & $257.5 \pm 6.1$ & 224 \\
\hline
$B^-$&$193 \pm 11$ & 163 \\
\hline
$B_s^0$&$253 \pm 8 \pm 7$ & 170 \\
\hline
$B_c$& $489\pm 5 \pm 3$~\cite{Chiu:2007bc} & 489 \\
\hline
\end{tabular}

\vspace*{1cm}

\def\arraystretch{1.25}
\caption{Decay constants $f_V$ of vector mesons with open flavor in MeV.}
\label{tab:5}
\begin{tabular}{|l|c|c|c|l|}
\hline
Meson &Data & \ Our \ \\ \hline
$\rho^+$ & $210.5 \pm 0.6$~\cite{Beringer:1900zz}& 216 \\ \hline
$D^\ast$&$245\pm 20^{+3}_{-2}$~\cite{Becirevic:1998ua}& 207 \\ \hline
$D_s^\ast$&$272\pm 16^{+3}_{-20}$~\cite{Aubin:2005ar}& 224 \\ \hline
$B^\ast$&$196\pm24^{+39}_{-2}$~\cite{Becirevic:1998ua}& 163 \\ \hline
$B^\ast_s$&$229\pm20^{+41}_{-16}$~\cite{Becirevic:1998ua}& 170 \\ \hline
\end{tabular}

\vspace*{1cm}

\def\arraystretch{1.25}
\caption{Decay constants $f_V$ of vector mesons with hidden flavor in MeV.}
\label{tab:6}
\begin{tabular}{|l|c|c|c|}
\hline
Meson &Data~\cite{Beringer:1900zz} & \ Our \ \\ \hline
$\rho^0$& 154.7 $\pm$ 0.7 & 153  \\ \hline
$\omega$ & 45.8 $\pm$ 0.8 & 51   \\ \hline
$\phi$   & 76   $\pm$ 1.2 & 72 \\ \hline
$J/\psi$ &277.6 $\pm$ 4   & 223  \\ \hline
$\Upsilon(1s)$& 238.5 $\pm$ 5.5 & 170 \\ \hline
\end{tabular}
\end{table}

\end{widetext}

\end{document}